# Monte Carlo Modeling and Design of Photon Energy Attenuation Layers (PALs) for 10-30x Quantum Yield Enhancement in Si-based Hard X-ray Detectors

Eldred Lee[1,2] [*], Michael R. James[2], Kaitlin M. Anagnost[1], Zhehui Wang[2], Eric R. Fossum[1], Jifeng Liu[1] [**]

[1] Thayer School of Engineering, Dartmouth College, Hanover, NH 03755, USA
[2] Los Alamos National Laboratory, Los Alamos, NM 87545, USA

**Abstract:**

High-energy (>20keV) X-ray photon detection at high quantum yield, high spatial resolution and short response time has long been an important area of study in physics. Scintillation is a prevalent method but limited in various ways. Directly detecting high-energy X-ray photons has been a challenge to this day, mainly due to low photon-to-photoelectron conversion efficiencies. Commercially available state-of-the-art Si direct detection products such as the Si charge-coupled device (CCD) are inefficient for >10keV photons. Here, we present Monte Carlo simulation results and analyses to introduce a highly effective yet simple high-energy X-ray detection concept with significantly enhanced photon-to-electron conversion efficiencies composed of two layers: a top high-Z photon energy attenuator layer (PAL) and a bottom Si detector. We use the principle of photon energy down conversion, where high-energy X-ray photon energies are attenuated down to ≤10keV via inelastic scattering suitable for efficient photoelectric absorption by Si. Our Monte Carlo simulation results demonstrate that 10-30x increase in quantum yield can be achieved using PbTe PAL on Si, potentially advancing high-resolution, high-efficiency X-ray detection using PAL-enhanced Si CMOS image sensors.

[*] Eldred.Lee.TH@dartmouth.edu or elee@lanl.gov

[**] Jifeng.Liu@dartmouth.edu

Since the discovery of X-rays by Roentgen, the continuous expansion of X-ray technology has transformed our society from materials science to biomedical applications. However, the capability gap for efficient and direct detection of high-energy X-ray photons in the 20-50keV range and beyond is challenging, which will prevent further advancements in rising fields of technology such as the generic platform of X-ray imaging sensor technology, quantum computing, and the next generation of synchrotron light source facilities [1-5]. Scintillator-based methods are widely used in these types of facilities and high-energy X-ray detection technologies; however, they have major limitations such as decay time response and light yield [6,7]. Furthermore, the spatial resolution of such method is limited due to the thickness of absorber materials and light propagation. There are also commercially available state-of-the-art X-ray detectors based on Si direct detection. For instance, the Si charge-coupled device (CCD) and CMOS image sensors, which are commercially available in many formats, are known to be efficient in X-ray photon energies ≤10keV and typically most efficient in the 100eV-10keV range [8]. However, beyond this energy, Si is known to be notoriously inefficient in detecting X-rays [8,9].

There have been considerable efforts to overcome these issues and limitations by many researchers. One example of such work is a two-layer high-energy X-ray detector structure comprising a cm-scale-area and mm-scale-thick metal buildup layer on a semiconductor thin-film to obtain high energy deposition and high dose delivery for radiation therapy [10,11]. But because this type of work could be unsuitable with small-form-factor commercial-off-the-shelf technologies such as CMOS image sensors (CIS) and quanta image sensors (QIS) as mm- and cm-scale metal could cause significant X-ray crosstalk to adjacent pixels that might be 20-50μm in pixel pitch [12]. Furthermore, the claim that has been made for the underlying principle of this prior work is that high-energy X-ray photons are converted to secondary electrons through Compton interaction in the top mm-scale metal buildup layer followed by the transport of these secondary electrons into the semiconductor thin-film layer for electron detection [10]. While this concept is intriguing, one questionable critical matter for our energy range of interest is that if the incident X-ray photon energies are relatively low compared to those intended for medical applications (MeV range), the low-energy secondary electrons (~1-10keV range) could never escape the thick metal buildup layer after absorbing an X-ray photon as the corresponding electron mean free paths are rather small (<1μm) in many materials [13]; therefore, the electrons could never make it to the bottom semiconductor thin-film layer for electron detection. Because of the relatively small travel ranges of electrons, even if the semiconductor thin-film layer for electron detection is decreased to μm-scale in terms of thickness, it is likely that the secondary electrons could still remain in the mm-scale-thick metal buildup layer.

Another class of work focuses on structured photocathode designs that depend on external photoemission, angle of X-ray photon absorption, and collection of electrons produced by photon interactions using an external electric field. This approach has demonstrated up to 5% quantum yield (QY), which is defined [14-16] as

$$QY = \frac{\text{\# of primary photoelectrons collected}}{\text{\# of incident photons}} \quad (1).$$

For almost 4 decades, QY from external photoemission has been roaming around 1-1.5%; therefore, the increase to 5% is commendable [16,17]. While the technique to reach the 5% QY is claimed to be suitable for higher X-ray photon energies at around 20-30keV, one should note that the X-ray photon absorption coefficient will decay rapidly with the increase of incident photon energies, which eventually will lead to lower QY at higher incident photon energies [8,9,18]. Therefore, this technique may not be extended to higher X-ray photon energy ranges of interest (i.e. >20keV).

Here, we present a new concept for direct detection of high-energy X-ray photons by using a high-Z thin film, the "photon attenuation layer" (PAL), to attenuate the incident photon energy below 10keV, thereby allowing more efficient absorption of down-converted X-ray photons by Si detectors underneath, or more generally, an "electron generation layer" (EGL). Our Monte Carlo simulation results and analyses using Monte Carlo N-Particle Software (MCNP6.2) demonstrates 10-30x QY enhancement in the X-ray photon energy range of 20-50keV. The conceptual design consists of two layers: a high-Z material PAL and a Si EGL, as shown in Fig. 1(a). The two types of primary interaction mechanisms in PAL-EGL are photon energy down conversion due to inelastic collisions, followed by photoelectric absorption [19] (Fig. 1(a)). Because X-ray absorption and scattering cross-sections increase as atomic numbers increase, high-Z materials are chosen for the PAL [20]. In PAL, photons lose energies via single and multiple inelastic collision events and eventually undergo notable red-shift from the incident X-ray wavelength, leading to an effective X-ray photon energy attenuation down to ≤10keV. This then allows Si to undergo photoelectric absorption with much higher absorption coefficient, thereby significantly improving the QY. Depending on the Si thickness, there could also be additional photon energy attenuation within Si prior to the photon-to-electron conversion. Cascade processes such as impact ionization, which could lead to further multiplication gain, occur following the photoelectric absorption because the average energy of X-ray excited primary photoelectrons is on the order of keV, while the average energy required to create an electron-hole pair (EHP) in Si is only 3.65eV [21-23]. In addition, it is expected that the PAL-EGL concept can also be integrated with CIS- or QIS-based devices. A schematic diagram of the expected CIS- or QIS-based device cross-section with PAL-EGL integration is shown in Fig. 1(b).

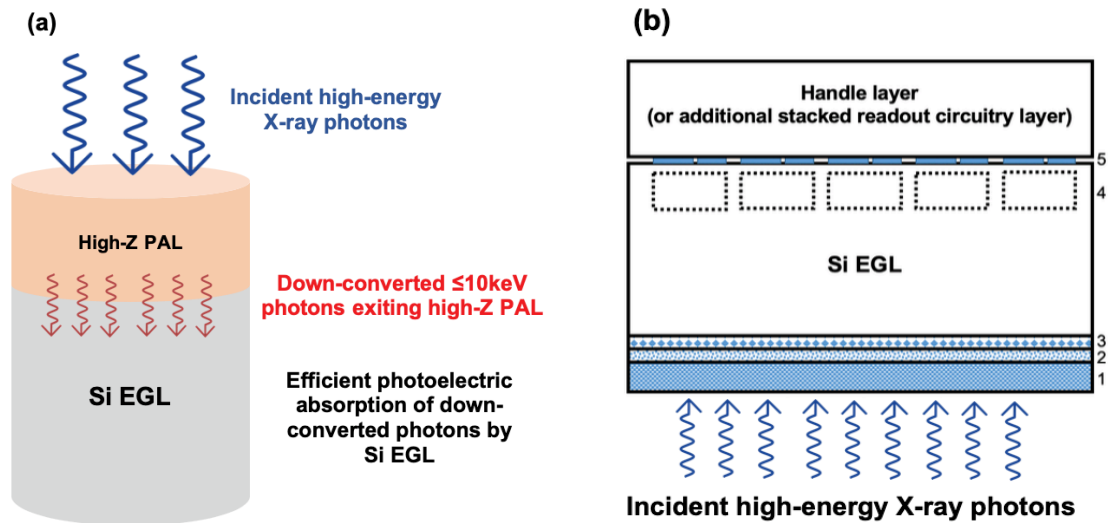

FIG. 1. (a) Schematics showing the mechanism of high-Z PAL-Si EGL detector design. High-energy X-ray photons (shown with blue arrows) are incident to the top of the high-Z PAL. Incident photons are down-converted (i.e. redshifted, as shown with red arrows) via inelastic scattering in the high-Z PAL and undergo efficient photoelectric absorption by Si. (b) Representation of the expected cross-section of backside illuminated CIS- or QIS-based device with PAL-EGL integration. 1 is high-Z PAL and 2 is a thin (<100 nm) backside passivation oxide (e.g. $SiO_2$) that can be added between PAL and Si for practicality but the overall QY should not be affected because down-converted X-ray photons can readily penetrate through the oxide layer. 3 is optional implants or epitaxial growth for surface pinning, 4 are pixelated carrier storage wells, and 5 is front side pixel readout circuitry.

While the overall underlying principle of photon energy down conversion could be somewhat similar to scintillator-based methods, it should be emphasized that this approach is distinctive in that the attenuated photons *still remain in the X-ray spectral regime* as opposed to the UV and visible regime. Unlike scintillators, the down conversion primarily relies on inelastic scattering with high-Z atoms, therefore no exotic and expensive bulk crystals (as in the case of scintillators) are needed for the PAL layers. In fact, the PAL layers can be polycrystalline or even amorphous thin films, which are much easier to fabricate than bulk crystal scintillators. The response time is also no longer limited by the optical spontaneous emission lifetime in scintillators, potentially allowing for ultrafast response since X-ray photon energy down conversion time via X-ray fluorescence and/or inelastic scattering is typically much shorter than the optical fluorescence time in scintillators. This conceptual design may also offer integration capabilities to Si CIS- or QIS-based devices for high resolution X-ray imaging.

For the purpose of this Letter, the thickness of the high-Z PAL was set as 1μm for most data to simply demonstrate and emphasize that a thin high-Z material layer can tremendously enhance high-energy X-ray photon energy attenuation, leading to efficient photoelectric absorption in Si. On the other hand, PAL thicknesses can be optimized corresponding to the thicknesses of Si and incident X-ray photon energies, which will be discussed towards the end of this Letter. Si layer thicknesses were chosen according to the typical range between CIS/QIS and commercial Si wafers. High-Z semiconductor materials such as CdTe, CdZnTe (CZT), and PbTe can be used as PAL materials due to their chemical stability and material availability for thin film solar cells or infrared detectors [24-27]. For this Letter, we have primarily explored PbTe because Pb has a higher atomic number than Cd. Furthermore, PbTe is also easier to fabricate than CdTe and CZT thin films [28].

To verify the concept of photon energy attenuation in thin-film PAL using high-Z semiconductor materials, energy distributions of the photons transmitted through the high-Z PAL have been modeled with MCNP6.2. The MCNP simulation was conducted on $10^5$ incident photons, and the transmitted photon energy histograms are plotted in 0.1keV bins. We confirm that the transmitted photons have a much lower energy than the incident ones, and a notable fraction will have energies <10keV to facilitate absorption by Si as shown in Figs. 2(a) and 2(b) for 20 and 30keV incident X-ray photons after transmitting through a 1μm PbTe PAL, respectively. These figures do not show the energy range of 0-1keV since there is a default artificial photon energy cutoff around 1keV in our MCNP simulation and important effects for scattering leading to lower energies are not yet included in MCNP6.2 photon transport methods, resulting in photon energies below the cutoff to not be calculated [29,30]. This cutoff tends to underestimate the photon absorption in Si because the mass attenuation coefficients of Si decreases with photon energy, as shown in Figs. 2(a) and 2(b) to demonstrate overlaps with the down-converted X-ray photon spectra [13,31]. From Fig. 2, we find that the photon energies are down-converted mainly into two distinctive regimes after transmitting through 1μm-thick PbTe PAL:

(1) A nearly continuous low energy spectrum at 1-5keV, where Si has large mass attenuation coefficients for efficient absorption. This regime is induced by multiple inelastic scattering of incident photons.

(2) Sharp and discrete energy peaks corresponding either to the characteristic X-ray emissions of Pb or Te atoms [32-34], or to the incident photon energy subtracted by the energy losses from the absorption edges of Pb or Te. Full tables that list the origins of a majority of the significant peaks are provided in the Supplemental Material. This regime is induced by photons that have only experienced from one to a few inelastic scattering events.

For the case of 20keV incident photons, several major peaks in regime (2) are still located at <10keV, which can be effectively absorbed by Si. Therefore, both regimes (1) and (2) contribute significantly to enhanced X-ray absorption in Si in this case. As the incident photon energy increases above 30keV, most of these characteristic X-ray peaks are located at >10keV (Fig. 2b), and efficient absorption of down-converted photons at 1-5keV in regime (1) become the dominant mechanism of QY enhancement for Si detectors.

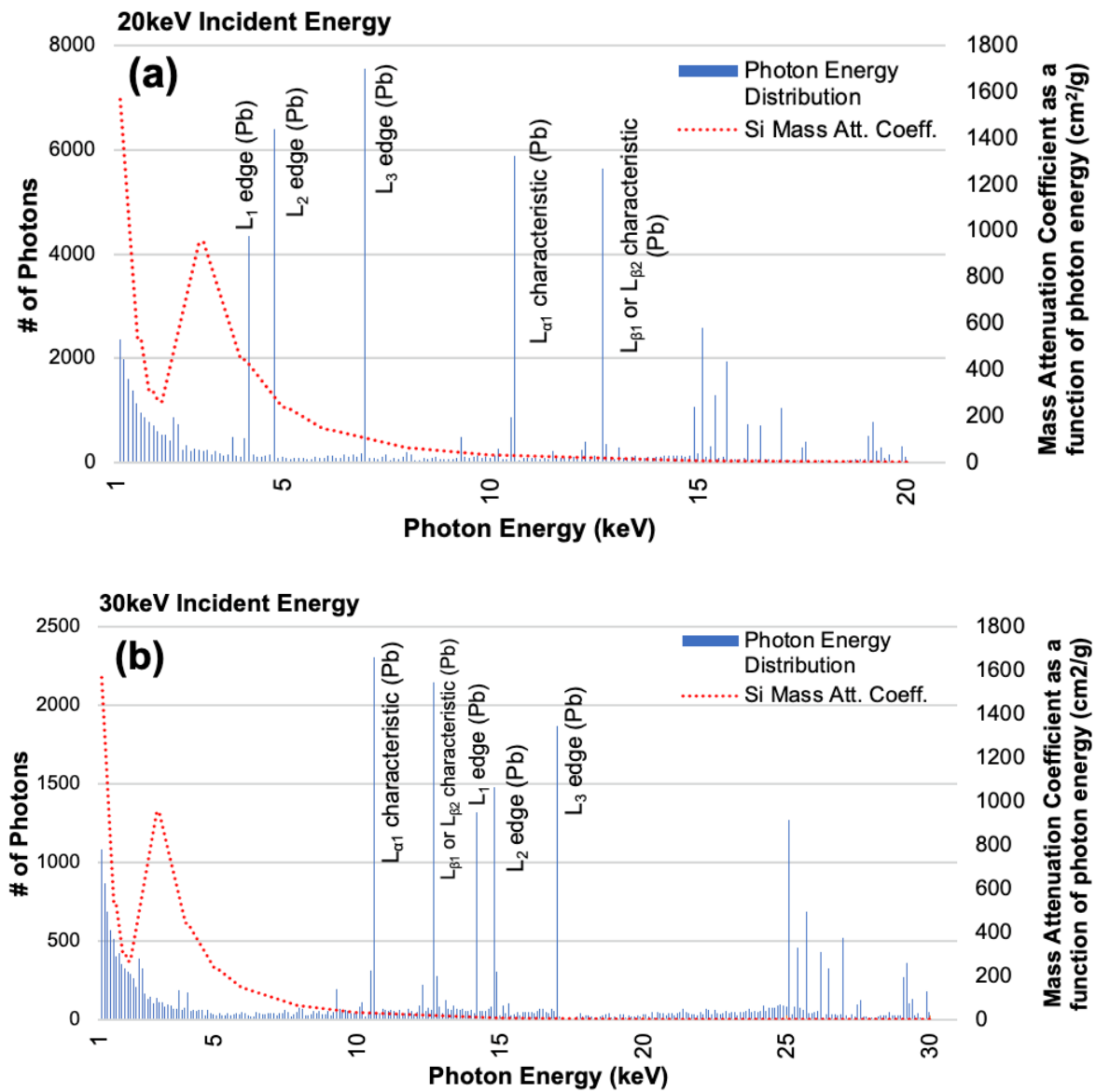

FIG. 2. X-ray photon energy distribution histograms after transmitting through a 1μm-thick PbTe PAL layer for (a) 20keV and (b) 30keV incident photons. The total number of incident photons is $10^5$, and the histograms are plotted using 0.1keV energy bins. Each significant discrete peak in the histogram is identified/labeled with either characteristic X-ray emission of Pb/Te atoms, or incident photon energy subtracted by the energy losses from the absorption edges of Pb/Te. Mass attenuation coefficients of Si as a function of photon energy are also shown in both plots. This mass attenuation coefficient spectrum indicates that X-ray photon energies need to be attenuated below 10keV for efficient absorption by Si.

Further including 5μm, 50μm, and 200μm-thick Si layers in the MCNP simulation with $10^5$ incident photons, Figs. 3(a) and 3(b) compare the QY of Si hard X-ray detectors with and without PAL as a function of incident photon energy based on the definition in Equation (1). Figs. 3(a) and 3(b) demonstrate that incorporating a 1μm-thick PAL layer can effectively increase the QY of Si detectors by 10-30x depending on the incident photon energy and the Si thickness. Even though the mass attenuation coefficient of Si decreases with photon energy (Fig. 2), Fig. 3(b) shows that the QY enhancement contributed by PbTe PAL actually increases with the incident X-ray photon energy. In other words, devices with PAL show much less QY degradation at higher X-ray photon energies. This feature is especially helpful for high-energy X-ray detection. For 5μm, 50μm, and 200μm Si, the QY with 1μm PbTe PAL (solid lines) ranges between 6.54% and 33.48% for 20keV photons. Remarkably, even the thinnest 5μm Si with PAL demonstrates ~2x higher QY than the thickest 200μm Si without PAL. Furthermore, QYs with PALs are all higher than the ~5% QY at 7.5keV incident X-ray photon energy as demonstrated by state-of-the-art photoemission X-ray detectors [13].

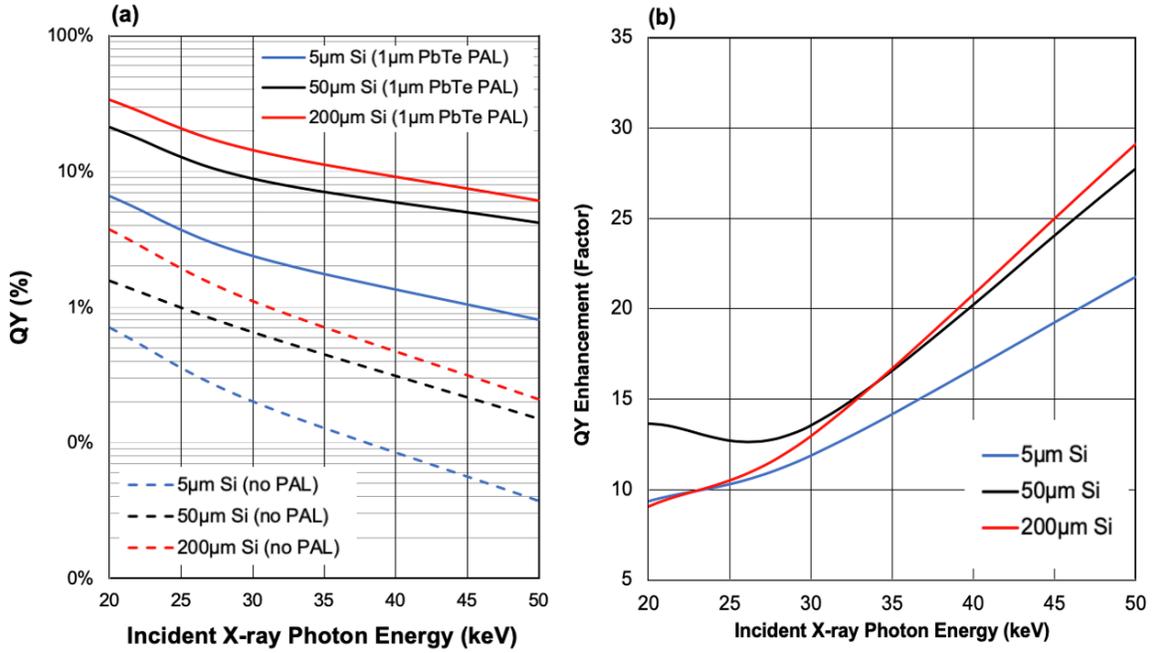

FIG 3. (a) QY in Si as a function of incident photon energy with 1μm PbTe PAL (solid lines) and without 1μm PbTe PAL (dashed lines) and with 5μm, 50μm, and 200μm Si. (b) QY enhancement (factor) vs. incident photon energy for the three different Si thicknesses to represent that 1μm PbTe PAL can remarkably increase the QY as opposed to having no PAL.

The average energies of primary photoelectrons in Si from Fig. 3(a) solid lines are found to be in the keV range; therefore, multiplication gain processes due to regenerative actions should take place. As previously stated, impact ionization processes to promote electrons in the valence band to the conduction band can further provide significant multiplication gain to the number of electrons that will be generated within Si. The number of primary photoelectrons upon X-ray excitation associated with the QY and the approximate total # of electrons after impact ionization are shown in Table I. To approximate the number of electrons post-multiplication gain processes, the following equation can be taken into consideration:

$$\text{\# of electrons in Si after multiplication gain}$$

$$= \frac{\text{avg. primary photoelectron energy in Si [eV]} * \text{\# of primary photoelectrons in Si EGL}}{\frac{3.65 eV}{EHP}}. \quad (2)$$

In Eq. (2), the average energies and the number of the primary photoelectrons generated in Si prior to additional multiplication gain processes can be determined using MCNP6.2 for devices with 1μm PbTe PAL (i.e. corresponding to the solid lines in Fig. 3(a)). The product of these two should be divided by 3.65eV, the aforementioned average energy required to generate an EHP in Si [3,24-26], to approximate the post-multiplication gain of the ultimate total number of electrons in Si. Table I shows the comparison between the number of X-ray excited primary photoelectrons before multiplication gain (which were used to determine the QY in Fig. 3(a)) and the approximate number of electrons post-multiplication gains using Eq. (2). All cases lead to an approximate post-multiplication gains of $10^3$-$10^4$.

Table I. Comparison between the number of photoelectrons (before impact ionization/multiplication) per $10^5$ incident X-ray photons and ultimate # of electrons after impact ionization/multiplication in 5μm, 50μm, and 200μm Si with 1μm PbTe PAL. Average electron energies in Si are still in the keV; therefore, further multiplication gain processes can be undergone to provide at least three orders of magnitude gain as shown on the far-right column of the table. As expected, thicker Si leads to higher number of electrons, which will lead to higher QY.

| Incident Energy (keV) | Si Thickness (um) | Avg. Electron Energy in Si (keV) | # of Primary Photoelectrons Pre-multiplication Gain [Fig. 3(a) Solid Lines] | Approx. Ultimate # of Electrons Post-multiplication Gain |
|---|---|---|---|---|
| 20 | 5 | 4.83 | 6535 | 8.65E+06 |
| 30 | 5 | 6.79 | 2373 | 4.41E+06 |
| 50 | 5 | 11.00 | 805 | 2.43E+06 |
| 20 | 50 | 4.39 | 21123 | 2.54E+07 |
| 30 | 50 | 5.75 | 8799 | 1.39E+07 |
| 50 | 50 | 8.45 | 4162 | 9.64E+06 |
| 20 | 200 | 4.24 | 33478 | 3.89E+07 |
| 30 | 200 | 5.43 | 14260 | 2.12E+07 |
| 50 | 200 | 7.73 | 6156 | 1.30E+07 |

It was previously mentioned that optimal thicknesses of PAL corresponding to incident X-ray photon energies and Si thicknesses can be determined. To confirm this statement, different thicknesses of PbTe PAL were placed on top of 5μm and 200μm Si at 20keV, 30keV, and 50keV incident X-ray photon energies to demonstrate that PAL thicknesses can be optimized according to different incident energies and Si thicknesses. In Fig. 4(a), with 5μm Si and the three different incident energies, a peak between 1μm and 1.5μm PbTe PAL is seen, indicating that 1μm-1.5μm PbTe PAL will result in highest QY with 5μm Si at the corresponding incident energies. In Fig. 4(b), with 200μm Si and the three different incident energies, a peak between 0.5μm and 0.75μm PbTe PAL is seen; therefore, 0.5μm-0.75μm PbTe PAL will result in the highest QY with 200μm Si. Notably, a QY approaching 40% can be achieved for 20keV incident photons, and 16% for 35keV incident photons. These results further confirm the potentials of PAL layers for integration with Si-based high-energy X-ray detectors.

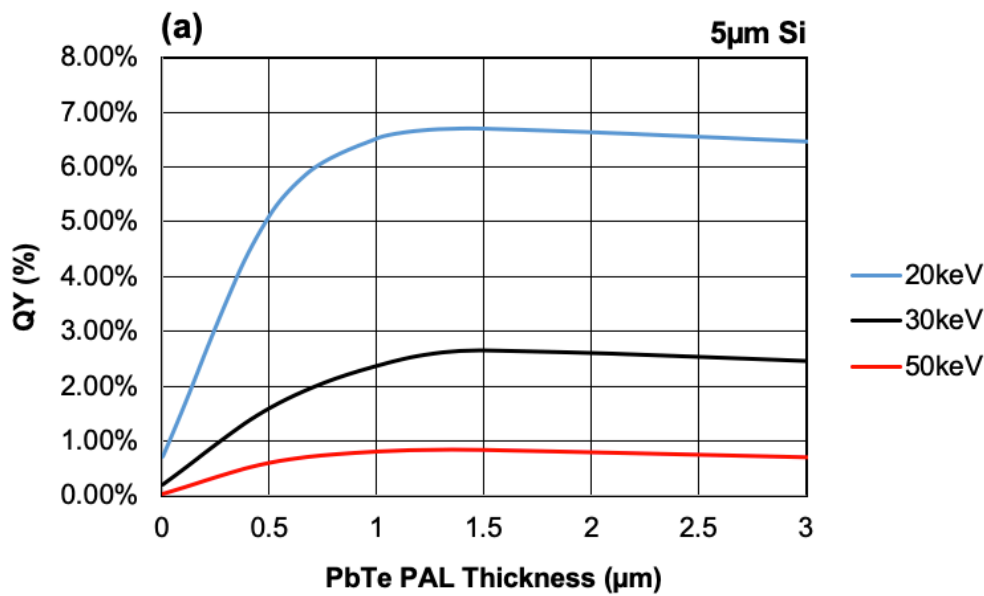

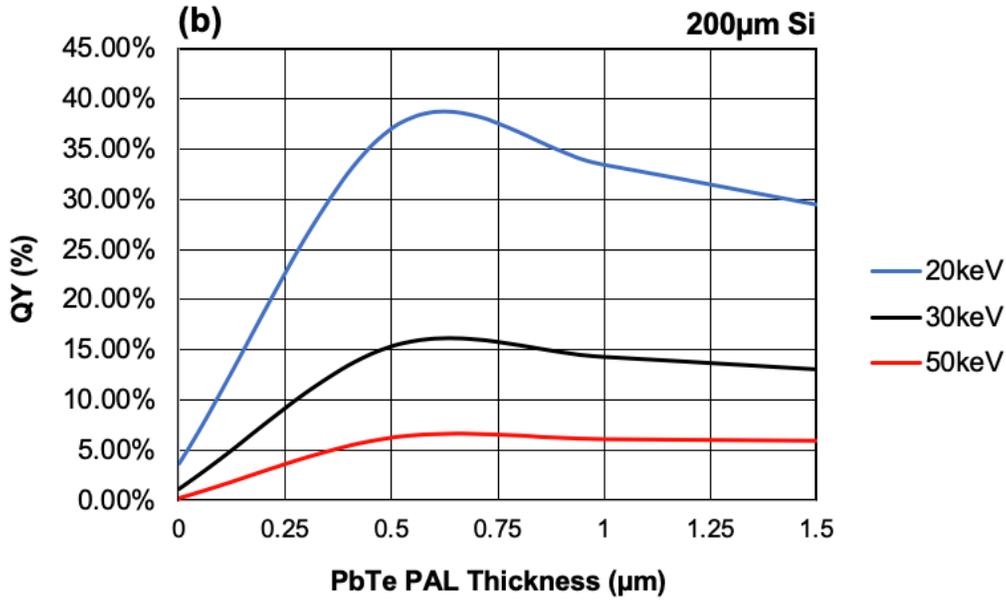

FIG. 4.  QY of (a) 5μm Si and (b) Si with different thicknesses of PbTe PAL at 20keV, 30keV, and 50keV incident X-ray photon energies. Optimal PbTe PAL thickness is seen between 1μm and 1.5μm for Si, and between 0.5μm and 0.75μm for 200μm Si.

In conclusion, we introduced a new high-energy X-ray direct detection concept capable of enhancing the QY by 10-30x for Si-based X-ray detectors using high-Z PAL, with a great potential to well surpass the performance of state-of-the-art X-ray detectors based on Si CCD or photocathodes. This simple yet highly effective device structure and its underlying principle of X-ray photon energy down-conversion have the potential to transform X-ray detection, e.g. spatial resolution and response time. Additional schemes of PAL layer material optimization are possible. Furthermore, with the capability of monolithic integration with Si CIS, the PAL-enhanced image sensors can also pave the way towards a wide field-of-view X-ray camera designs for synchrotron and X-ray free electron laser light source applications [3,4]. The modeling in this work will guide future experimental verification towards high-resolution, high-efficiency X-ray detection using PAL-enhanced Si CIS.

**Acknowledgements**

This study was supported by P-25 Subatomic Physics Group at Los Alamos National Laboratory under subcontract number 537679 and under basic agreement number 537992 with The Trustees of Dartmouth College. It was also supported by the United States Department of Energy National Nuclear Security Administration Laboratory Residency Graduate Fellowship (DOE NNSA LRGF) under award number DE-NA0003864. Los Alamos National Laboratory is managed by Triad National Security, LLC for the United States Department of Energy's NNSA.

# Monte Carlo Modeling and Design of Photon Energy Attenuation Layers (PALs) for 10-30x Efficiency Enhancement in Si-based Hard X-ray Detectors – Supplemental Material

Eldred Lee[1,2]*, Michael R. James[2], Kaitlin M. Anagnost[1], Zhehui Wang[2], Eric R. Fossum[1], Jifeng Liu[1]**

[1] Thayer School of Engineering, Dartmouth College, Hanover, NH 03755, USA
[2] Los Alamos National Laboratory, Los Alamos, NM 87545, USA

**1.** Table of peaks in 0.1keV bin photon energy distribution plot representing energies corresponding to energy losses from incident energies (edges) or characteristic X-ray energies: 1um PbTe PAL at 20keV incident energy. Energy losses from incident are the differences between incident energies and peak energies.

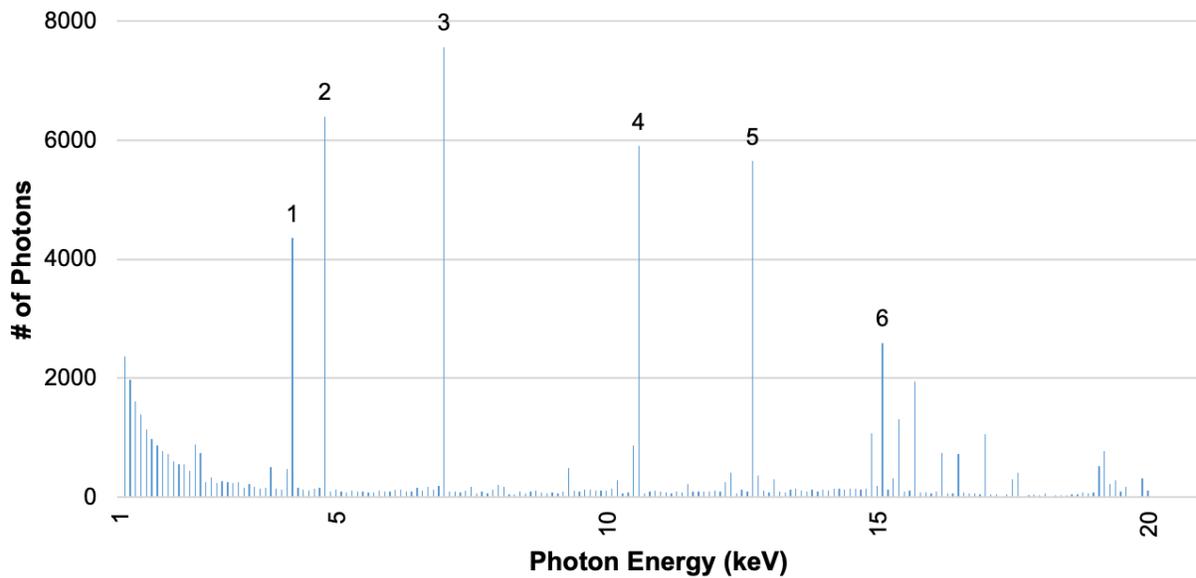

| Peak # | Peak Energy (keV) | Energy Loss from Incident (keV) | Transition Corresponding to Loss | Characteristic Corresponding to Peak |
|---|---|---|---|---|
| 6 | 15.1 | 4.9 | $L_1$ edge (Te) or M (Pb) | -- |
| 5 | 12.7 | 7.3 | -- | $L_{\beta 1}$ or $L_{\beta 2}$ (Pb) |
| 4 | 10.6 | 9.4 | -- | $L_{\alpha 1}$ (Pb) |
| 3 | 7.0 | 13.0 | $L_3$ edge (Pb) | -- |
| 2 | 4.8 | 15.2 | $L_2$ edge (Pb) | -- |
| 1 | 4.2 | 15.8 | $L_1$ edge (Pb) | -- |

**2.** Table of peaks in 0.1keV bin photon energy distribution plot representing transition energies corresponding to energy losses from incident energies (edges) or characteristic X-ray energies: 1um PbTe PAL at 30keV incident energy.

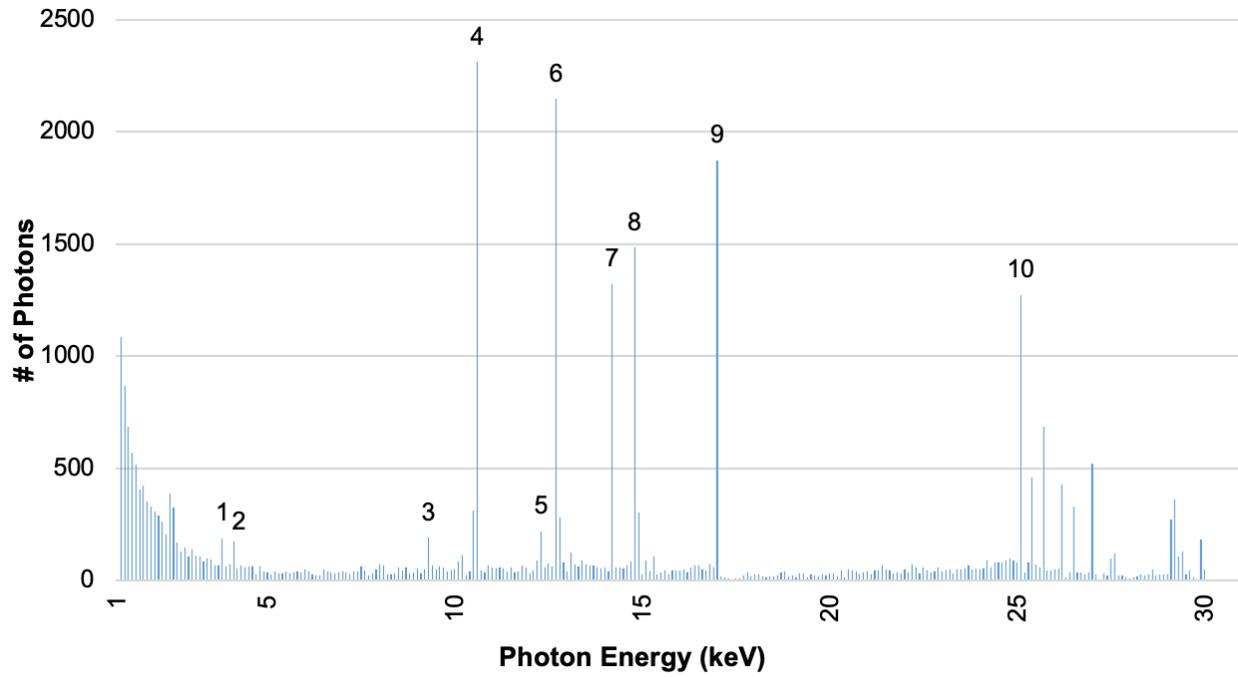

| Peak # | Peak Energy (keV) | Energy Loss from Incident (keV) | Transition Corresponding to Loss | Characteristic Corresponding to Peak |
|---|---|---|---|---|
| 10 | 25.1 | 4.9 | $L_1$ edge (Te) or M (Pb) | -- |
| 9 | 17.0 | 13.0 | $L_3$ edge (Pb) | -- |
| 8 | 14.8 | 15.2 | $L_2$ edge (Pb) | -- |
| 7 | 14.2 | 15.8 | $L_1$ edge (Pb) | -- |
| 6 | 12.7 | 17.3 | -- | $L_{\beta 1}$ or $L_{\beta 2}$ (Pb) |
| 5 | 12.3 | 17.7 | -- | $L_{\beta 4}$ (Pb) |
| 4 | 10.6 | 19.4 | -- | $L_{\alpha 1}$ (Pb) |
| 3 | 9.3 | 20.7 | -- | $L\ell$ or $L\tau$ (Pb) |
| 2 | 4.1 | 25.9 | -- | $L_{\beta 3}$ (Te) |
| 1 | 3.8 | 46.2 | -- | $L_{\beta 9}$ (Te) |